\pdfoutput=1
\documentclass{PoS}

\usepackage{dsfont}
\usepackage{amssymb}
\usepackage{slashed}
\usepackage{tabulary}
\usepackage{bbold}
\usepackage{bm}
\usepackage{amsmath}
\usepackage{subfigure}

\title{Renormalization of the momentum density on the lattice using shifted boundary conditions}

\ShortTitle{Renormalization of the momentum density on the lattice using shifted boundary conditions}

\author{\speaker{Daniel Robaina}\\
       PRISMA Cluster of Excellence, Institut f\"ur Kernphysik, \newline Johannes Gutenberg-Universit\"at Mainz, D-55099 Mainz, Germany\\
       E-mail: \email{robaina@kph.uni-mainz.de}}

\author{Harvey B. Meyer\\
        PRISMA Cluster of Excellence, Institut f\"ur Kernphysik and Helmholtz Institut Mainz, \newline Johannes Gutenberg-Universit\"at Mainz, D-55099 Mainz, Germany\\
        E-mail: \email{meyerh@kph.uni-mainz.de}}

\abstract{In order to extract transport quantities from energy-momentum-tensor (EMT) correlators in Lattice QCD there is a strong need for a non-perturbative renormalization of these operators. This is due to the fact that the lattice regularization explicitly breaks translational invariance, invalidating the non-renormalization-theorem. Here we present a non-perturbative calculation of the renormalization constant of the off-diagonal components of the EMT in SU(3) pure gauge theory using lattices with shifted boundary conditions. This allows us to induce a non-zero momentum in the system controlled by the shift parameter and to determine the normalization of the momentum density operator.}

\FullConference{31st International Symposium on Lattice Field Theory LATTICE 2013\\
		 July 29 - August 3, 2013\\
		 Mainz, Germany}

\begin{document}

\section{Introduction}
In the modern approach to study static and dynamic properties of a strongly interacting quantum system like the Quark-Gluon-Plasma (QGP), calculating the transport coefficients plays an important role. These ``low energy constants'' of hydrodynamics, viewed as an effective field theory, can be extracted from spectral functions calculated in the underlying theory via the famous Kubo-formulas (see for instance \cite{Kubo}). As a prominent example we can mention the shear viscosity $\eta$ which parametrizes how efficiently the momentum of a layer of fluid diffuses in the direction orthogonal to the layer. It should be obvious that transport of energy and momentum is governed by energy-momentum-tensor-correlators.\\
\indent Due to the lack of continuous Poincar\'e invariance induced by the lattice regularization, the discretized Energy-momentum-tensor (EMT) is only conserved up to cutoff effects invalidating the general argument that \emph{``correlation functions among conserved currents are already renormalized."}
Hence, on the lattice we have to further investigate the renormalization pattern of $T^{\mu \nu}$ in order to have a well defined continuum limit of its correlation functions and be able to safely extract transport coefficients.\\
\indent In this talk we will address the problem of calculating the renormalization constant $Z_T$ of the momentum densitiy $T^{0k}$ for SU(3) pure Yang-Mills theory using shifted boundary conditions \cite{HB3}. The work is organized as follows: in section 2 we give a precise formulation of the problem under study and explain the renormalization pattern of the EMT on the lattice. In section 3 the concept of shifted boundaries is explained and some important results are quoted. In section 4 we apply the strategy and explain the precise method for computing $Z_T$ non-perturbatively. Finally, in section 5 we present our results and draw some conclusions.

\section{Renormalization pattern of $T_{\mu \nu}$}
In the continuum, the pure SU(3) EMT can be splitted into a traceless part and the trace itself,
\begin{equation}
T_{\mu \nu} = \theta_{\mu \nu} + \frac{1}{4}\delta_{\mu \nu} \theta,
\end{equation}
where
\begin{equation}
\theta_{\mu \nu} = \frac{1}{4} \delta_{\mu \nu} F^a_{\rho \sigma}F^a_{\rho \sigma} - F^a_{\mu \alpha} F^a_{\nu \alpha} \;\;\;\; \rm{and} \;\;\;\; \theta = \emph{T}_{\mu \mu}.
\end{equation}
On the lattice, we discretize $\theta_{\mu \nu}$ as written in Eq. (2.2) by using the ``clover" form of the field strength tensor as defined in \cite{Luescher}.

In order to proceed with the renormalization of the momentum densitiy $T^{0k}$ we need to know the transformation properties of this operator under the symmetry group of the lattice: the hypercubic group denoted by H(4).
The operators $T_{\mu \nu}$ split into irreducible representations of H(4). In particular, $T^{0k}=\theta^{0k}$ belongs to a six dimensional representation along with the other off-diagonal components of $T^{\mu \nu}$ \cite{Goeck}. Since there are no other gauge invariant operators of equal or less dimension, $T^{0k}$ renormalizes multiplicatively
\begin{equation}
T^{\textrm{R}}_{\mu \nu} = Z_T T_{\mu \nu} \;\;\;\;\;\;\;\;   \mu \neq \nu .
\end{equation}
Our aim will be to determine in a non perturbative way the renormalization constant $Z_T = Z_T(g^2_0)$. Since in the continuum the momentum density is a conserved current, this renormalization constant does not depend on any external scale but only on the bare coupling $g^2_0$.
In the classical field theory, $Z_T=1$ and due to asymptotic freedom, we expect
\begin{equation}
\lim_{g^2_0 \rightarrow 0} Z_T = 1 .
\end{equation}
The $\mathcal{O}(g^2_0)$ correction has been computed in Lattice Perturbation Theory in \cite{Carraciolo}. 
\section{Shifted boundary conditions}
Shifted boundaries were investigated by Giusti and Meyer in \cite{HB1,HB2,HB3}. The starting point is the partition function
\begin{equation}
Z(L_0, \xi) = \text{Tr}\{e^{-L_0(\hat{H}-i\xi \hat{P})}\}.
\end{equation}
In the Euclidean path integral formalism, the phase $e^{iL_0\xi \cdot p}$ is encoded in the shifted boundary conditions of the fields, (see Fig. 1)
\begin{equation}
\phi(L_0, x) = \phi (0, x+L_0\xi).
\end{equation}
Based on Lorentz symmetry considerations the authors demonstrated a relationship between a ``shifted" system at inverse temperature $L_0$ and an unshifted system at inverse temperature $L_0\sqrt{1+\xi^2}$ where $L_0$ is the temporal extension and $\xi$ the shift parameter.\footnote{It should be noticed that periodic boundary conditions are kept in spatial directions.}
The free energies
\begin{equation}
f(L_0, \xi)= -\frac{1}{L_0 V} \log Z(L_0, \xi)
\end{equation}
of both systems are exactly equal when the thermodynamic limit is taken,
\begin{equation}
f(L_0, \xi) = f(L_0\sqrt{1+\xi^2}, 0) .
\end{equation}
\begin{figure}[ht]
\begin{center}
\def\svgwidth{9cm}
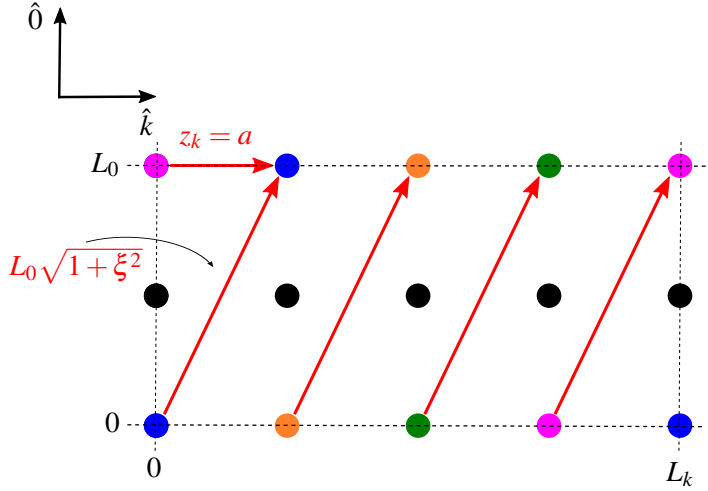
\caption{Shifted lattice with renormalized shift parameter ${\xi}_k = \frac{z_k}{L_0}$ for a $4 \times 2$ lattice.}
\end{center}
\end{figure}
\noindent The finite volume corrections are exponentially suppressed in the lightest screening mass of the theory. Having proven this relation, one can derive non-trivial Ward Identities connecting both systems by taking derivatives with respect to $L_0$ and $\xi_k$ \cite{HB1}. In this way, one shows that the entropy of a system at inverse temperature $\beta=L_0\sqrt{1+\xi^2}$ can be obtained by measuring the momentum density at nonzero shift
\begin{equation}
s=-\frac{Z_T L_0 (1+\xi^2)^{3/2}}{\xi_k}\left<T^{0k}\right>_{\xi} .
\end{equation}
We will use the entropy as the key observable to extract the renormalization constant as it will be explained in the next section.

In addition to the possibility of being able to extract thermodynamic potentials like entropy or specific heat among others, shifted boundaries have also other applications. Since the shift parameter is a ratio of two physical lengths, it does not need any renormalization. Again, because of Lorentz invariance, the temperature can only depend on the combination $L_0\sqrt{1+\xi^2}$. Therefore, we end up with far more freedom to set the temperature because of the additional parameters $\xi_1, \xi_2$ and $\xi_3$. In particular, there is no need to change the lattice spacing nor the temporal extension in order to perform scans in temperature. This represents a technical advantage because the ultraviolet behaviour of the theory does not get changed.\footnote{This means that renormalization and improvement coefficients are kept constant when varying $\xi^2$.} This can be of special importance when one is near a phase transition and wants to vary the temperature in much smaller steps. We encourage interested readers to learn more about shifted boundaries in \cite{HB1,HB2,HB3}.

\section{Method}
Before moving on to the extraction of $Z_T$ we performed consistency checks on the free theory where this constant takes a value of 1. In \cite{HB1} the authors calculated in perturbation theory the expectation value of the momentum density in the free lattice theory for different values of the shift. Via Eq. (3.5) they extracted the ratio of the entropy at finite lattice spacing with respect to the continuum value
\begin{equation}
\frac{\left(s\beta^3\right)}{\left(s_{SB}\beta^3\right)} = 1 + c_1(a/L_0)
\end{equation}
where $s_{SB}\beta^3 = 32\pi^2/45$ for SU(3) and $c_1$ is a coefficient that encodes the lattice discretization effects and depends only on $(a/L_0)^2$. It has to  approach zero when the true continuum limit is taken.
In order to reproduce this ratio we had to simulate at very small values of $g^2_0$. This leads to some complications. On the one hand the critical slowdown of the algorithm makes it harder to maintain autocorrelation effects under control. On the other hand, finite size effects are of order $\sim e^{-g^2_0 L/\beta}$ where $\beta$ is the inverse temperature as described in the previous section. This requires that one chooses a very large aspect ratio $L/\beta$.
\subsection{The case $L_0/a=2$ and $\xi=(1,0,0)$}
Having decided on this simplest case we measured the entropy via Eq. (3.5) for different values of the coupling with increasing volumina in order to extrapolate to zero coupling as shown in Fig. (2a). The intercept takes a value of
\begin{equation}
\frac{(s\beta^3)}{(s_{SB}\beta^3)} = 1.018(6)
\end{equation}
which compared to the analytic value calculated in lattice perturbation theory taken from \cite{HB1}
\begin{equation}
\frac{(s\beta^3)}{(s_{SB}\beta^3)} = 1.026
\end{equation}
gives a standard deviation of $\sim 1.2\sigma$, indicating that errors are under control.


\subsection{Computation of $Z_T(g^2_0)$}
In order to extract the relevant renormalization constant we need some already renormalized quantity to compare with. For this purpose, following \cite{HB2}, one can define a \emph{Generating Function associated with the Momentum Distribution},
\begin{equation}
K(L_0, \boldsymbol{\xi}) = - \log \frac{Z(L_0, \boldsymbol{\xi})}{Z(L_0)} = -\log \frac{\text{Tr}\{e^{-L_0(\hat{H}-i\boldsymbol{\xi} \boldsymbol{\hat{P}})}\}}{\text{Tr}\{e^{-L_0 \hat{H}}\}},
\end{equation}
which is nothing but a ratio of two partition functions: in the nominator the one with shifted boundary conditions and in the denominator the ordinary partition function with periodic boundary conditions. It turns out that this is a RGI quantity and therefore has a finite and universal continuum limit. Any derived quantity that one can compute by taking derivatives with respect to this functional is already renormalized, too. In particular,
\begin{equation}
\frac{\partial K(L_0, \boldsymbol{\xi})}{\partial \boldsymbol{\xi_{k}}} = - L_0 Z_T \left<\int d^3x\, T_{0k}(x)\right>_{\boldsymbol{\xi}}.
\end{equation}
How to systematically compute ratios of partition functions was first investigated in \cite{dellaM1,dellaM2}. This motivated the study of the contribution to any field theoretical partition function due to states with a given set of quantum numbers. In our case, $K(L_0, \boldsymbol{\xi})$ is related by Fourier transform to the relative contribution of states with momentum $\bold{p}$ (see \cite{HB3} for a more precise explanation).

If we stick to the case of $\bold{\xi}=(\xi_1,0,0)$ it can be shown that the cumulants of this functional are related to thermodynamic potentials \cite{HB1}. For example, the entropy of the system at finite lattice spacing is defined by
\begin{equation}
\left(\frac{s}{T^3}\right)_R = \frac{\partial^2}{\partial \xi^2_1} \left.\frac{K(L_0, \bold{\xi})}{T^3 L^3}\right|_{\bold{\xi}=0} = \frac{2 K(L_0, \bold{\xi})}{|\bold{\xi}|^2T^3L^3}
\end{equation}
where an $R$ subscript indicates that this is a renormalized quantity.
Following Eq. (4.5), one can come to the conclusion that by measuring the expectation value of the momentum density along the $x$-direction at nonzero shift and taking the first derivative with respect to $\xi_1$ one ends up with an unrenormalized prediction for $(s/T^3)_0$ that can be written as the traditional definition of the derivative:
\begin{equation}
\left(\frac{s}{T^3}\right)_0 = \frac{\partial}{\partial \xi_1} \left.\frac{\left<T_{01}\right>_{\xi}}{T^4}\right|_{\xi=0} =\lim_{\xi_1 \rightarrow 0} \left[\frac{\left<T_{01}\right>_{\boldsymbol{\xi}}}{T^4 \xi_1} \right].
\end{equation} 
The last term cancels out because of symmetry considerations. Since $\xi_1$ is not a continuous parameter that can be tuned we have to stick to rational values and let them be close to zero where the derivative needs to be evaluated. The best choice is to take $\xi_1= a/L_0$ to get more accurate results for increasing number of points in the temporal direction. In addition, since $\left<T_{01}\right>_{\boldsymbol{\xi}}$ is an odd function in $\xi_1$ our errors are of $\mathcal{O}(a/L_0)^2$.

Combining the last two equations it is clear that
\begin{equation}
Z_T = \frac{\left(\frac{s}{T^3}\right)_R}{\left(\frac{s}{T^3}\right)_0}
\end{equation}
which is the master equation for calculating the renormalization constant of interest.

\section{Results \& Conclusions}
After performing the fit to the data our final result is (see Fig. (2b) and Table 1)
\begin{equation}
Z_T = \frac{1+0.1368g^2_0 + 0.1858g^4_0}{1-0.1332g^2_0}.
\end{equation}
The absolute error is $\sim 0.008$ when $g^2_0 \in [0.8, 1.0]$. In order to connect with perturbative results we used as an input for our fit function the one-loop approximation that was calculated in \cite{Carraciolo}. It should be noticed that the good agreement between the three different data sets corresponding to different temperatures is an indication that we have $\mathcal{O}(aT)^2$ errors under control since $Z_T$ should not depend on any external scale.
\section*{Acknowledgments}
We thank Leonardo Giusti and Stefano Capitani for useful discussions and valuable comments. The simulations were performed on the dedicated QCD platform ``Nikola'' at the Institute for Nuclear Physics, University of Mainz. This work was supported by the \emph{Center for Computational Sciences in Mainz} as part of the Rhineland-Palatinate Research Initiative and by the DFG grant ME3622/2-1 \emph{Static and dynamic properties of QCD at finite temperature}.

\begin{table}[ht]
\begin{center}
\begin{tabular}{ccccccc}
Lat        &$6/g^2_0$&$L_0/a$&$L/a$ & $\left(\frac{s}{T^3}\right)_{R}$ & $ \left(\frac{s}{T^3}\right)_{0}$ &$\textcolor{red}{Z_T}$ \\[0.1cm]
\hline\\[-0.4cm]
${\rm A}_2$   &$6.024$& $5$&$16$ &$4.98(4)$ & 3.27(3)&\textcolor{red}{1.52(2)}\\[0.2cm]
${\rm A}_3$   &$6.137$& $6$&$18$ &$4.88(6)$ & 3.24(4)&\textcolor{red}{1.50(3)}\\[0.2cm]
${\rm A}_4$   &$6.337$& $8$&$24$ &$5.12(19)$ & 3.28(6)&\textcolor{red}{1.56(6)}\\[0.2cm]
${\rm A}_5$   &$6.507$&$10$&$30$&$4.9(3)$ & 3.32(9)&\textcolor{red}{1.47(10)}\\[0.10cm] %
\hline\\[-0.4cm]
${\rm B}_2$   &$6.747$& $5$&$16$ &$6.53(6)$ & 4.53(2)&\textcolor{red}{1.44(2)}\\[0.2cm]
${\rm B}_3$   &$6.883$& $6$&$18$ &$6.40(6) $ & 4.53(2)&\textcolor{red}{1.41(2)}\\[0.2cm]
${\rm B}_4$   &$7.135$& $8$&$24$ &$6.42(20)$& 4.54(2)&\textcolor{red}{1.41(4)}\\[0.2cm]
${\rm B}_5$   &$7.325$&$10$&$30$&$6.1(3)$& 4.55(4)&\textcolor{red}{1.33(7)}\\[0.10cm]%
\hline\\[-0.4cm]
${\rm C}_2$   &$7.426$& $5$&$20$ &$7.13(8)$ & 5.11(3)&\textcolor{red}{1.39(2)}\\[0.2cm]
${\rm C}_3$   &$7.584$& $6$&$24$ &$6.94(12)$ & 5.07(3)&\textcolor{red}{1.36(3)}\\[0.05cm]
\hline
\end{tabular}
\end{center}
\caption{$A_\#$ corresponds to $T=1.5T_C$, $B_\#$ correspond to $T=4.1T_C$ and $C_\#$ corresponds to $T=9.1T_C$ .  The values of the column $\left(\frac{s}{T^3}\right)_{R}$ were taken from \cite{HB3}.}
\end{table}

\begin{figure}[h]
\hspace{-2cm}
\subfigure[]{
\input{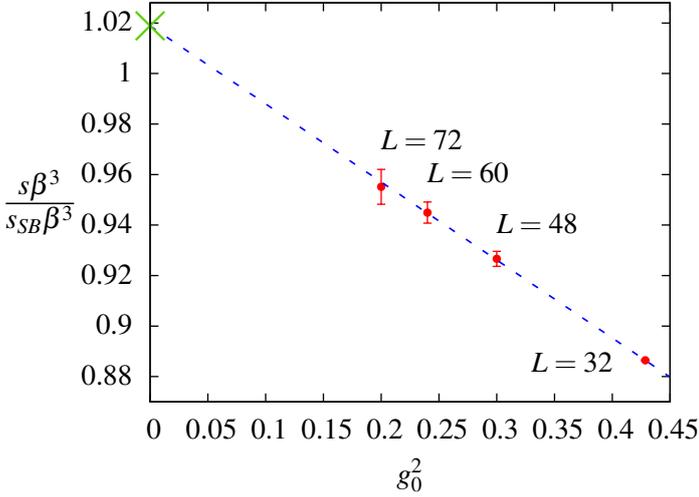}
}
\qquad
\hspace{-1.1cm}
\subfigure[]{
\input{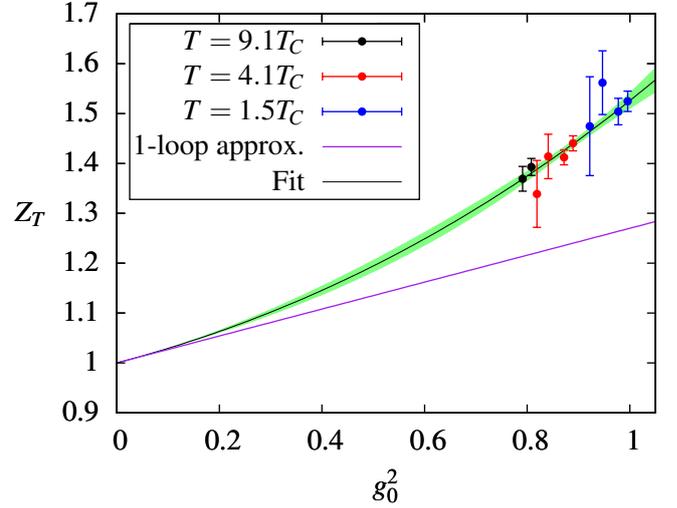}
 }
\caption{(a) Extrapolation to the free lattice theory at finite shift $\xi=(1,0,0)$ and $L_0/a=2$. (b) $Z_T$ as a function of $g^2_0$ for three different values of temperature.}
\end{figure}

\end{document}